\documentstyle[12pt]{article}
\newcommand{\bec}{\begin{center}}
\newcommand{\ec}{\end{center}}
\newcommand{\bee}{\begin{equation}}
\newcommand{\ee}{\end{equation}}
\newfont{\blackboard}{msbm10 scaled\magstep2}
\newcommand{\Z}{\mbox{\blackboard\symbol{"5A}}}
\newcommand{\C}{\mbox{\blackboard\symbol{'103}}}
\newcommand{\R}{\mbox{\blackboard\symbol{'122}}}
\newcommand{\N}{\mbox{\blackboard\symbol{'116}}}

\begin{document}
\large
\begin{titlepage}
\bec
{\Large\bf  D-branes and twisted K-theory  \\}
\vspace*{15mm}
{\bf Yuri Malyuta \\}
\vspace*{10mm}
{\it Institute for Nuclear Research\\
National Academy of
Sciences of Ukraine\\
03022 Kiev, Ukraine\\}
e-mail: interdep@kinr.kiev.ua\\
\vspace*{35mm}
{\bf Abstract\\}
\end{center}
\vspace*{1mm}
Topological charges of the $D6$-brane
in the presence of a Neveu-Schwarz B-field
are computed by methods of twisted $K$-theory.
\vspace*{1cm}
\hspace*{2.3cm} Extensions of $C^{*}$-algebras.
\end{titlepage}
\section {\bf Introduction}
\hspace*{6mm}$D$-branes are topological solitons,
which charges are described by Grothendieck K-groups
\cite{1.}. We begin our consideration with
instantons \cite{2.} as they ideologically remind
$D$-branes.

	The self-dual equation for the $SU(2)$ instanton
has the form
$\hspace*{6cm}F_{\mu\nu}=F^{*}_{\mu\nu}\ ,\hfill (1.1)$\\
where Yang-Mills strengths and potentials are
defined as follows
\[F_{\mu\nu}=\partial_{\mu}A_{\nu}-\partial_{\nu}A_{\mu}
+[A_{\mu}, A_{\nu}], \ \ \
F^{*}_{\mu\nu}=\frac{1}{2}\varepsilon_{\mu\nu\lambda\kappa}
F_{\lambda\kappa},
 \ \ \ \varepsilon_{0123}=1,\]
\[F_{\mu\nu}=\frac{1}{2i}\sigma^{a}F_{\mu\nu}^{a}(x) , 
\ \ \
A_{\mu}=\frac{1}{2i}\sigma^{a}A^{a}_{\mu}(x) ,\]
$\sigma^{a}/2$ are generators of the $SU(2)$
group, $x=(x_{0}, x_{1}, x_{2}, x_{3})$
are coordinates in Euclidean space.

	't Hooft \cite{3.} has found the
$5k$ parameter solution of the equation (1.1)
\[A_{\mu}=\frac{1}{2i}{\eta_{\mu\nu}}
\partial_{\nu}ln\phi \ ,\]
where \hspace*{1.7cm} $\eta_{\mu\nu}=
(\varepsilon_{oa\mu\nu}+
\delta_{a\mu}\delta_{o\nu}-
\delta_{a\nu}\delta_{o\mu})\sigma^{a} \ , $
\[\phi=1+\sum\limits_{i=1}^{k}
\frac{\lambda^{2}_{i}}{(x_{\nu}-y_{\nu i})^{2}} \ , \ \ \
k=\frac{1}{16\pi^{2}}
\int d^{4}x Tr(F_{\mu\nu}F^{*}_{\mu\nu}) \ , \]
$\lambda_{i}$ and $y_{\nu i}$ are parameters, $k$ 
is the topological charge.

	It was shown in the works of Atiyah,
Hitchin, Drinfeld and Manin \cite{4.,5.}
that dynamics of the equation (1.1) is encoded
in the principal bundle
\[\hspace*{5cm}\begin{tabular}{cccc}
$SU(2)$\hspace*{-0.2cm}&$\rightarrow$ &
\hspace*{-0.3cm}$P$ & \\
&    &\hspace*{-2.5mm}$\downarrow$ &
\hspace*{4cm} (1.2) \\
 &   &\hspace*{-2.5mm}$S^{4}$ &    \\
\end{tabular}\]
%\vspace*{5mm}\\
and in the associated vector bundle

\[\hspace*{5.8cm}\begin{tabular}{cccc}
$\C^{2}$\hspace*{-0.2cm}&$\rightarrow$ &
\hspace*{-0.3cm}$P$ & \\
&    &\hspace*{-2.5mm}$\downarrow$ &
\hspace*{4cm} (1.3) \\
 &   &\hspace*{-2.5mm}$S^{4}$ &    \\
\end{tabular}\]
%\vspace*{5mm}\\
The potential $A_{\mu}$
is a connection of the vector bundle (1.3),
therefore it is possible to calculate it 
by methods of algebraic geometry.
For this purpose following the techniques
of Horrocks and Barth \cite{6.}, it is
necessary to calculate the orthonormalized 
basis of sections $Y^{p}$
of the vector bundle (1.3). Then the potential 
$A_{\mu}$ will be equal to 
\[A_{\mu}^{pq}=(Y^{p}, \partial_{\mu}Y^{q}) \ .\]
The topological charge  $k$ is an element
of the homotopy group
\[\pi_{3}(SU(2))=\Z \ ,\]
characterizing the principal bundle (1.2).

	In the next sections we shall apply the 
algebro-geometric approach to the description
of $D$-branes.
\section{Twisted K-theory}
\hspace*{6mm}Let us consider the following
principal bundles describing $D$-branes 
\cite{7.}
\bec
\[\hspace*{5cm}\begin{tabular}{cccc}
\hspace*{-1.5cm}$U(n)$&\hspace*{-4mm}
$\rightarrow$ &
\hspace*{-0.3cm}$P_{H}$ &   \\
&    &\hspace*{-2.5mm}$\downarrow$ & 
\hspace*{5.7cm}(2.1) \\
 &   &\hspace*{-1.5mm}$X$ &
\end{tabular}\]
\ec
\vspace*{7mm}
\bec
\[\hspace*{5.15cm}\begin{tabular}{cccc}
\hspace*{-1.5cm}$SU(n)/\Z_{n}$&
\hspace*{-4mm}$\rightarrow$ &
\hspace*{-0.3cm}$P_{H}$ &   \\
&    &\hspace*{-2.5mm}$\downarrow $ 
&\hspace*{4.85cm}   (2.2)\\
 &   &\hspace*{-1.5mm}$X$ &
\end{tabular}\]
\ec
\vspace*{7mm}
\bec
\[\hspace*{5cm}\begin{tabular}{cccc}
\hspace*{-1.5cm}
$\lim\limits_{n\rightarrow\infty}SU(n)/
\Z_{n}$&\hspace*{-4mm}$\rightarrow$ &
\hspace*{-0.3cm}$P_{H}$ &   \\
&    &\hspace*{-2.5mm}$\downarrow $ & 
\hspace*{3.8cm} (2.3)\\
 &   &\hspace*{-1.5mm}$X$ &
\end{tabular}\]
\ec
\vspace*{9mm}
where $X$ is a compact manifold.\\
This bundles are characterized by the
Dixmier-Douady invariant \cite{8.}
\[[H]\in H^{3}(X, \Z) \ , \]
which represents the strength of a
Neveu-Schwarz $B$-field \cite{9.}
interacting with $D$-branes
\[H_{\mu\nu\rho}=\partial_{\mu}B_{\nu\rho}+
\partial_{\nu}B_{\rho\mu}+\partial_{\rho}B_{\mu\nu}\ .\]
For the bundle (2.1)\\
\hspace*{2cm}$[H]=0 \ ,$ \ \ 
\ \ id est \ \ \ $H_{\mu\nu\lambda}=0 \ , 
\ \ \
B_{\mu\nu}=0 \ ;$\\
for the bundle (2.2)\\
\hspace*{2cm}
$n[H]=0 \ ,$\ \ \ id est \ \ \ $H_{\mu\nu\lambda}=0 \ , 
\ \ \
B_{\mu\nu}\neq 0 \ ;$ \\
for the bundle (2.3)\\
\hspace*{2cm}
$[H]\neq 0 \ ,$ \ \ \ \  id est \ \ \
$H_{\mu\nu\lambda}\neq 0 \ , \ \ \
B_{\mu\nu}\neq 0 \ .$

	The vector bundles associated
with the principal bundles (2.1), (2.2), (2.3)
are given by \cite{7.}\\
\vspace*{-3mm}\\
$E_{H}=P_{H}\times \C^{n} \ ,$\ \ \ where 
\ \ \ $Aut(\C^{n})=U(n)\ ;$
\hfill (2.4)\\
$E_{H}=P_{H}\times M_{n}(\C) \ ,$\ \ \ where \ \ \
$Aut(M_{n}(\C))=SU(n)/\Z_{n}\ ;$
\hfill (2.5)\\
$E_{H}=P_{H}\times {\cal K} \ ,$\ \ \ where \ \ \
$Aut({\cal K})=\lim\limits_{n\rightarrow\infty}SU(n)/
\Z_{n}\ ;$
\hfill (2.6)\\
\vspace*{-3mm}\\
$M_{n}(\C)$ is the algebra of  $n\times n$ matrices, 
$\cal K$ is the algebra of compact operators.\\
The space of sections of the vector bundle
(2.4) is the algebra of continuous
functions $C_{0}(X)$;\\
the space of sections of the vector bundle 
(2.5) is the $C^{*}$-algebra 
$C_{0}(X, E_{H})$ , 
which is Morita-equivalent
to the Azumaya algebra \cite{10.};\\
the space of sections of the vector bundle (2.6)
is the Rosenberg $C^{*}$-algebra 
$C_{0}(X, E_{H})$ \cite{11.}.

	Let us consider the short 
exact sequences \cite{12.}
\[\hspace*{3cm}
0 \rightarrow
{\cal K}  \rightarrow
{\cal B}  \rightarrow
C_{0}(X , E_{H})  \rightarrow
0 \hspace*{3cm} (2.7)\]
\[\hspace*{3cm}
0 \rightarrow
{\cal K}  \rightarrow
{\cal B^{'}}  \rightarrow
C_{0}(X , E_{H})  \rightarrow
0 \hspace*{2.9cm} (2.8)\]
determining extensions of the algebra
$C_{0}(X, E_{H})$ by the algebra ${\cal K}$.\\
If in the commutative diagrams
\vspace*{-0.2cm}\\
\[0 \rightarrow {\cal K} \rightarrow{\cal B} 
\rightarrow C_{0}(X , E_{H})
\rightarrow 0\]
\[\hspace*{-0.4cm}\downarrow \hspace*{1cm}
\downarrow \hspace*{1.6cm}
{\downarrow}{\tau_{1}}\]
\[0 \rightarrow {\cal K} \rightarrow B({\cal H}) 
\stackrel{\pi}{ \rightarrow}
{\cal Q(H)}\rightarrow 0\]
\vspace*{0.3cm}
\[0 \rightarrow {\cal K} \rightarrow {\cal B^{'}} 
\rightarrow C_{0}(X , E_{H})
\rightarrow 0\]
\[\hspace*{-0.5cm}\downarrow \hspace*{1cm} 
\downarrow \hspace*{1.6cm}
{\downarrow}{\tau_{2}}\]
\[0 \rightarrow {\cal K} \rightarrow B({\cal H}) 
\stackrel{\pi}{ \rightarrow}
{\cal Q(H)}\rightarrow 0\]
\vspace*{-0.5cm}\\
(where $B({\cal H})$ is the algebra
of bounded operators on a Hilbert space ${\cal H}$, 
${\cal Q(H)}$ denotes the Calkin algebra)
morphisms are related by
\[\tau_{2}(s)=\pi (U)\tau_{1}(s)\pi (U)^{*}\]
(s $\in$ $C_{0}(X, E_{H})$, $U$ 
is a unitary operator on ${\cal H}$),
then two extensions (2.7) and (2.8)
are unitarily equivalent.

	Let $Ext(X, [H])$
denote the set of unitary equivalence
classes of extensions of $C_{0}(X , E_{H})$
by ${\cal K}$ modulo splitting
extensions. Then twisted K-groups are
defined as follows \cite{12.}
\[K_{j}(X, [H])=Ext(X\times \R^{j},\ p^{*}_{1}[H]), \ \ \
j \in \N \ ,\]
where $p_{1} : X\times \R^{j} \rightarrow X$
denotes projection onto the first factor.\\
The Bott periodicity theorem asserts:
\[K_{j}(X, [H])=\left\{
\begin{array}{rl}
K_{0}(X, [H]) & \mbox{for even } j\ ,\\
K_{1}(X, [H]) & \mbox{for odd } j\ .
\end{array}\right. \]
\section{Computation of twisted $K$-groups}
\hspace*{6mm}A short exact sequence
of $C^{*}$-algebras
\[0 \rightarrow
{\cal A}  \rightarrow
{\cal B}  \rightarrow
C_{0}(X , E_{H})  \rightarrow
0 \]
induces a hexagon exact sequence of
twisted K-groups \cite{12.}
\bec
\begin{tabular}{ccccccc}
&&\hspace*{-6.3cm}${K}_{0}(X, [H])$
&\hspace*{-5.4cm}$\rightarrow$
&\hspace*{-6.1cm}${K}_{0}({\cal B})$&\hspace*{-6.8cm}
$\rightarrow$
&\hspace*{-6.1cm}${K}_{0}({\cal A})$\\
\hspace*{3cm}${\uparrow}$
&&\hspace*{0cm} &&
\hspace*{0cm}${\downarrow}$&&  \hspace*{0.6cm} (3.1)\\
\hspace*{3cm}${K}_{1}({\cal A})$
&$\leftarrow$&${K}_{1}({\cal B})$
&\hspace*{-0cm}$\leftarrow$
&\hspace*{-0cm}${K}_{1}(X, [H])$&&\\
\end{tabular}
\ec
The hexagon sequence (3.1) enables us to
compute twisted K-groups.

	Let $X = S^{3}$. We will compute 
twisted groups $K_{0}(S^{3}, n[H])$ and 
$K_{1}(S^{3}, n[H])$ for the case
$H_{\mu\nu\lambda}=0 \ ,\ \ B_{\mu\nu}\neq 0$~.
There is the short exact sequence
\[0\rightarrow C_{0}(S^{2})\otimes {\cal K}
\rightarrow
C_{0}(U_{1})\otimes {\cal K}
 \ \oplus \ C_{0}(U_{2})\otimes {\cal K}
\rightarrow \]
\[\hspace*{8cm}\rightarrow
 C_{0}(S^{3}, E_{n[H]})
\rightarrow 0\]
(where $\{U_{1}, U_{2}\}$ 
is the open cover of $S^{3}$ by hemispheres)\ , \\
which induces the hexagon exact sequence of
K-groups
\bec
\hspace*{-2.9cm}\begin{tabular}{ccccccc}
\hspace*{3cm}${K}_{0}(S^{3}, n[H])$
&\hspace*{0cm}$\rightarrow$
&\hspace*{0cm}${K}_{0}(U_{1})\oplus K_{0}(U_{2})$&
\hspace*{0cm}
$\rightarrow$
&\hspace*{0cm}${K}_{0}(S^{2})$&&\\
\hspace*{3cm}${\uparrow}$
&&\hspace*{0cm} &&
\hspace*{0cm}${\downarrow}$&&\hspace*{-0.5cm}(3.2)\\
\hspace*{3cm}${K}_{1}(S^{2})$
&$\leftarrow$&${K}_{1}(U_{1})\oplus K_{1}(U_{2})$
&\hspace*{-0cm}$\leftarrow$
&\hspace*{-0cm}${K}_{1}(S^{3}, n[H])$&&\\
\end{tabular}
\ec
\vspace*{5mm}
Since
\[K_{0}(S^{2})=\Z \ , \hspace*{7mm} 
K_{0}(U_{1})\oplus K_{0}(U_{2})=\Z \ ,\]
\[K_{1}(S^{2})=0 \ , \hspace*{7mm} 
K_{1}(U_{1})\oplus K_{1}(U_{2})=0 \ ,\]
the sequence (3.2) reduces to the exact
sequence
\[\hspace*{13.5mm}0\rightarrow
K_{0}(S^{3}, n[H])\rightarrow
\Z \rightarrow
\Z \rightarrow
K_{1}(S^{3}, n[H])\rightarrow
0 \hspace*{0.64cm} (3.3) \]
Reduction of the sequence (3.3) to the
exact sequence
\[0\rightarrow
\Z \rightarrow
\Z \rightarrow
\Z_{n} \rightarrow
0\]
yields the following result:
\[K_{0}(S^{3}, n[H])=0 \ , \hspace*{7mm} K_{1}(S^{3}, 
n[H])=\Z_{n} \ .\]

	The group $K_{1}(S^{3}, n[H])=\Z_{n}$ 
determines
topological charges of the $D6$-brane for the 
case $H_{\mu\nu\lambda}=0 , \ \ B_{\mu\nu}\neq 0 .$\\
For comparision we present the
$Dp$-brane spectrum \cite{1.}
for the case $H_{\mu\nu\lambda}=0 , 
\ \ B_{\mu\nu} = 0 $ :
\vspace*{-0.2cm}\\
\bec
\begin{tabular}{|c|c|c|c|c|c|c|c|c|c|c|c|}
\hline
\normalsize$Dp$ &\normalsize$D9$
&\normalsize$D8$ &\normalsize$D7$
&\normalsize$D6$ &\normalsize$D5$
&\normalsize$D4$ &\normalsize$D3$
&\normalsize$D2$ &\normalsize$D1$
&\normalsize$D0$ &\normalsize$D(-1)$
\\  \hline
\normalsize$S^{9-p}$&\normalsize$S^{0}$
&\normalsize$S^{1}$ &\normalsize$S^{2}$
&\normalsize$S^{3}$ &\normalsize$S^{4}$
&\normalsize$S^{5}$ &\normalsize$S^{6}$
&\normalsize$S^{7}$ &\normalsize$S^{8}$
&\normalsize$S^{9}$ &\normalsize$S^{10}$
\\ \hline
\normalsize${K}_{0}(S^{9-p})$
&\normalsize$\Z$ &\normalsize0
&\normalsize$\Z$ &\normalsize0
&\normalsize$\Z$ &\normalsize0
&\normalsize$\Z$ &\normalsize0
&\normalsize$\Z$ &\normalsize0
&\normalsize$\Z$  \\ \hline
\normalsize${K}_{1}(S^{9-p})$
&\normalsize0 &\normalsize$\Z$
&\normalsize0 &\normalsize$\Z$
&\normalsize0 &\normalsize$\Z$
&\normalsize0 &\normalsize$\Z$
&\normalsize0 &\normalsize$\Z$
&\normalsize0  \\ \hline
\end{tabular}\\
\ec
\vspace*{2cm}
{\bf Acknowledgement}\\
This material was presented in the lecture
given by the author at the
Institute of Mathematics (Kiev, Ukraine).
The author thanks audience for 
questions and comments. 


\newpage
\begin{thebibliography}{99}
\bibitem{1.}E. Witten, {\it D-branes and K-theory}, 
hep-th/9810188.
\bibitem{2.}E. Corrigan, {\it Self-dual solutions to
Euclidean Yang-Mills equations}, Phys. Reports
{\bf 49} (1979) 95.
\bibitem{3.}R. Jackiw, C. Nohl and C. Rebbi,
{\it Conformal properties of
pseudoparticle configurations},
Phys. Rev. {\bf D15} (1977) 1642.
\bibitem{4.} V.G. Drinfeld and Yu.I. Manin,
{\it Self-dual Yang-Mills fields on the sphere},
Funkt. Anal. Priloz. {\bf 12}, N2 (1978) 78.
\bibitem{5.}M.F. Atiyah, N.J. Hitchin,
V.G. Drinfeld and Yu.I. Manin,
{\it Construction of instantons},
Phys. Lett. {\bf 65A} (1978) 185.
\bibitem{6.}W. Barth, {\it Moduli of vector
bundles on the projective plane}, 
Inventiones math. {\bf 42} (1977) 63.
\bibitem{7.}P. Bouwknegt and V. Mathai,
{\it $D$-branes, $B$-fields and 
twisted $K$-theory}, hep-th/0002023.
\bibitem{8.}J. Dixmier and A. Douady,
{\it Champs continues d'espaces hilbertiens
at de $C^{*}$-algebres}, Bull. Soc. Math. 
France {\bf 91} (1963) 227.
\bibitem{9.}E. Witten, {\it Overview of
$K$-theory applied to strings}, hep-th/0007175.
\bibitem{10.}A. Kapustin, {\it $D$-branes in
a topologically nontrivial $B$-field},
hep-th/9909089.
\bibitem{11.}J. Rosenberg, {\it Continuous
trace algebras from the bundle
theoretic point of view}, Jour. Austr. 
Math. Soc. {\bf 47} (1989) 368.
\bibitem{12.}V. Mathai and I.M. Singer,
{\it Twisted $K$-homology theory,
twisted Ext-theory}, hep-th/0012046.
\end{thebibliography}
\end{document}